# Une approche sociotechnique pour le Knowledge Management (KM)

Leoncio Jiménez

Departamento de Computación e Informática
Universidad Católica del Maule, Talca, Chile
ljimenez@spock.ucm.cl

**Résumé.** Cet article présente un cadre sociotechnique pour le KM. Cette vision sociotechnique du KM permet : (1) d'écarter le KM d'un souci commercial ; (2) faire le clivage des différentes technologies du KM ; et (3) de s'interroger sur les paradigmes associés aux composants social et technique du KM. C'est précisément ce dernier point que cet article développe afin d'identifier les mécanismes génériques du KM. Plus précisément, l'aspect social est décrit à travers l'approche organisationnelle du KM, l'approche managériale du KM, et l'approche biologique du KM, alors que l'aspect technique est décrit à travers l'approche ingénierie des connaissances et compétences du KM. Ces approches nous conduisent aussi à donner un tableau comparatif entre ces visions organisationnelles, managériales et biologiques du KM.

## 1   Introduction

L'hypothèse que le KM peut être observé comme un *système sociotechnique*[1] (Coakes *et al.*, 2002) traduit un chemin de pensée plus proche du constructivisme que du positivisme. Plus précisément afin d'intégrer dans une même démarche l'interrelation permanente de deux composants, on peut dire que l'un est social (il incarne les besoins des individus) et l'autre est technique (il représente l'impératif technologique que l'on veut) du système vu en termes entrées/transformation/sorties. Si l'on détaille plus, on peut dire que la composante sociale est formée par deux sous composants interrelationés : domaine (activité ou business) et acteurs (sujet et autres). De même, la composante technique est formée aussi par deux sous groupes appelés : tâches (processus et données) et technologie[2] (méthodes et outils). La richesse de l'approche réside alors dans sa complexité[3]. En effet l'ensemble (social et technique) coexiste de façon harmonieuse avec les autres, mais dans différents niveaux systémiques en générant ainsi plusieurs niveaux explicatifs d'une même réalité.

---
[1] Il s'agit d'un concept développé au début des années 50 par de chercheurs britanniques, visant à impliquer les individus et les groupes dans l'organisation des tâches. Herbst (1974) parle de l'école Tavistock de Londres, pour décrire ce mouvement.
[2] La technologie est l'ensemble de méthodes, de procédures, d'équipements et même d'approches utilisées pour fournir un service ou produire un bien. (Gousty, 1998).
[3] La complexité fait preuve des relations de causalité circulaire entre l'aspect social et l'aspect technique, et donc, la complexité n'a rien avoir avec la taille du système, le nombre de relations, ou les phénomènes compliqués du système.

Dans cet article, nous verrons tout d'abord l'aspect social du KM, puis nous présenterons brièvement l'aspect technique du KM. Nous conclurons en analysant les perspectives de l'ingénierie des systèmes de connaissances pour le KM dans un cadre sociotechnique.

## 2 Aspect social du KM

### 2.1 Approche organisationnelle du KM

Historiquement, la connaissance dans le travail a été considérée comme un facteur décisif pour les entreprises (Ballay, 1997), c'est ainsi qu'au cours du temps l'on trouve comme un levier d'avantage productif dans une économie de production (1930), concurrentiel ou compétitif dans une économie de service (1960), coopératif dans une économie globalisée (1990), et collective dans une économie du savoir (2000). D'où l'émergence de nouveaux systèmes d'organisation du travail et mode de management, autour de la connaissance, tels que : knowledge worker (1967), knowledge society (1969), learning organization (1990), systems thinking (1990), actionable knowledge (1996), knowledge-creating company (1995), information ecology (1997), information age (1997), knowledge-based economy (1997), corporate knowledge (2000), corporate longitude (2000), knowledge-based assets (2000), social responsibility in the information age (2005), etc.

L'approche organisationnelle du KM trouve ses racines dans le concept *knowledge-creating company* (connaissance créatrice organisationnelle) proposé par Nonaka et Takeuchi (1995) comme un facteur critique d'innovation et un levier d'avantage compétitif de l'entreprise. Ces auteurs indiquent que « l'entreprise ne "traite" pas seulement de la connaissance mais la "créée" aussi ». Dans ce même esprit non simonien Varela (1989) affirme « la métaphore populaire désignant le cerveau comme une machine de traitement de l'information n'est pas seulement ambiguë, elle est totalement fausse ». Dans ce contexte, une entreprise ne peut pas créer de la connaissance sans êtres vivants, car ils sont le moteur de la connaissance créatrice organisationnelle[4].

#### 2.1.1 Paradigme du ballon de rugby

L'équipe est à la base de la connaissance créatrice organisationnelle de l'entreprise (Nonaka et Takeuchi, 1995). Dans l'entreprise la connaissance existe dans un domaine individuel et collectif au travers de la convergence de deux dualités. Premièrement, la dualité sujet/objet traduit le fait que le sujet conscient (individu, groupe, entreprise) créée de la connaissance en s'impliquant lui-même dans l'objet (espace de travail, environnement, le travail et ses outils de production,…). Ceci forme la dimension épistémologique de la connaissance. Cette dimension a été visualisée à travers des travaux de Polanyi, développée dans ses livres, *Personal Knowledge* (publié en 1958) et *The Tacit Dimension* (publié en 1966), sur le champ phi-losophique de la "connaissance tacite" (*savoir-faire*). Pour Polanyi « les êtres humains acquièrent la connaissance en créant et organisant activement leurs propres expériences » (Nonaka et Takeuchi, 1995). Deuxièmement, la dualité sujet/autres traduit le fait que le sujet conscient créée de la connaissance dans la relation avec les autres (individu, groupe, entreprise). Ceci forme la dimension ontologique de la connaissance. Les travaux de Barnard sur la "connaissance comportementale" (*savoir-être*) développé dans *The Functions of the Executive* (publié en 1938), ont contribué à la formation de cette dimension.

---

[4] Certains auteurs Prax (2000), Boughzala et Ermine (2004) parlent davantage de "management" des connaissances et non pas de "gestion" des connaissances afin d'attirer l'attention dans l'aspect M du KM, c'est-à-dire l'aspect humain de la connaissance.

Nonaka et Takeuchi (1995) distinguent la connaissance tacite et explicite. La connaissance tacite est attachée à la conscience du sujet (niveau individuel) et du groupe (niveau collectif) à travers des idées, métaphores, créances, concepts, hypothèses, modèles mentaux, analogies, etc. La connaissance explicite est attachée à un objet (rapport, document, email, modèle, maquette, plan, etc.). Ainsi, la connaissance individuelle (tacite ou explicite) est au niveau individuel, la connaissance collective (tacite ou explicite) est au niveau du groupe, et la connaissance organisationnelle (tacite ou explicite) est au niveau de l'entreprise.

Un "ballon de rugby" permet de symboliser la capacité de l'entreprise de créer de nouvelles connaissances et de la valoriser sous forme de nouveaux produits ou services. Nonaka et Takeuchi (1995) affirment « la capacité d'une entreprise considérée dans son ensemble, de créer de nouvelles connaissances, de les disséminer au sein de l'organisation et de leur faire prendre corps dans les différents produits, services du système ». La création de connaissances n'est pas un fait isolé sinon qu'elle est liée à l'innovation continue et à l'avantage compétitif durable pour l'entreprise.

Le ballon (partie gauche figure 1) prend forme grâce à un processus de conversion de connaissances. Il s'agit d'un processus de causalité circulaire entre la dimension épistémologique (tacite, explicite) et ontologique (individu, groupe, entreprise) de la connaissance, à partir de quatre mécanismes génériques. De gauche à droite : *socialisation* : conversion de connaissance individuelle tacite à connaissance collective et organisationnelle tacite, la connaissance est créée à travers le partage de l'expérience par apprentissage organisationnel (observation, réflexion, construction de sens, implication personnelle, engagement individuel, etc.) ; *extériorisation* : conversion de connaissance individuelle tacite à connaissance collective et organisationnelle explicite, la connaissance est crée à travers l'explicitation (formalisation, énonciation) de la connaissance individuelle tacite. Recours au langage ou support écrit pour communiquer idées, concepts, analogies, métaphores, hypothèses, modèles mentaux, etc. ; *combinaison* : conversion de connaissance individuelle explicite à connaissance collective et organisationnelle explicite, la connaissance est créée à travers la mise en commun de la connaissance explicite à travers des réunions, de changements d'informations, de données, etc. ; *internalisation* : conversion de connaissance collective et organisationnelle explicite à connaissance collective et organisationnelle tacite, la connaissance dans l'entreprise est créée à travers la réflexion à partir des modèles mentaux partage, savoir-faire technique, etc. entre ses membres. Plus loin nous caractérisons ces mécanismes en termes d'apprentissage organisationnel.

Le ballon avance (partie droite figure 1) entre la dimension épistémologique et ontologique sous forme de spirale ascendante[5]. La connaissance tacite et explicite est disséminée entre les acteurs (individu, groupe, entreprise) à tous les niveaux de l'entreprise (stratégique, tactique, opérationnel).

---

[5] Le terme "spirale ascendante" indique un mouvement du ballon de rugby entre l'individu, le groupe et l'entreprise sous forme d'un processus de conversion (extériorisation, combinaison, internalisation, socialisation) entre la connaissance tacite et la connaissance explicite.

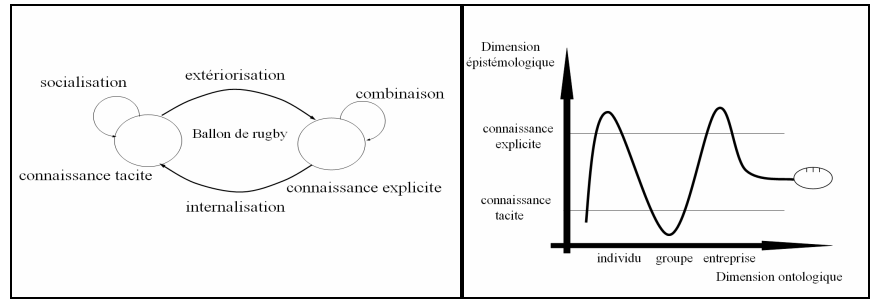

FIG. 1 – *Paradigme du ballon de rugby.*

L'apprentissage organisationnel prend forme dans la dimension épistémologique par une sorte de *savoir-faire* qui transforme la connaissance tacite en connaissance explicite (et vice versa), et d'autre part, dans la dimension ontologique comme un *savoir-être* entre l'individu, le groupe, et l'entreprise. Le KM est vu comme une dynamique de l'entreprise apprenante. Ingham commente dans l'introduction de l'édition française du livre de Nonaka et Takeuchi (1995), que « les processus d'apprentissage concernent les "savoir quoi faire" et les "savoir pourquoi faire" il s'agira d'apprendre "comment faire" et le résultat prendra généralement la forme d'un "savoir-faire". Mais ils pourront aussi entraîner une modification d'un comportement et avoir trait alors à un "savoir être" ».

L'approche organisationnelle du KM relie alors la connaissance à l'innovation (produits ou services), à l'avantage (productive, concurrentiel ou compétitif, coopératif, collectif) et à l'apprentissage organisationnel (dynamique de l'entreprise apprenante).

Gérer les connaissances de l'entreprise, caractérise les mécanismes de création de connaissances nouvelles et d'apprentissage organisationnel par la mise en place d'un processus de causalité circulaire parmi quatre états, dans le but de créer (connaissance créatrice organisationnelle), disséminer (individu, groupe, entreprise) et valoriser (nouveaux produits ou services) les connaissances de l'entreprise, ces états sont : *Socialisation* : *savoir* ou *savoir technique* pour exprimer la connaissance (tacite ou explicite) créée par le raisonnement de l'acteur (individu, groupe, entreprise), liés aux phénomènes d'intelligence humaine (individu, groupe), d'intelligence économique (entreprise), veille, etc. ; *Extériorisation* : *savoir-faire collectif* pour exprimer la connaissance organisationnelle ou collective, organisée comme un tout dans un système de connaissances et matérialisée par l'innovation (produits ou services) capable de produire un avantage compétitif durable pour l'entreprise ; *Combinaison* : *savoir-faire* ou *savoir-faire technique* pour exprimer la connaissance (tacite ou explicite) créée dans l'action (l'apprentissage) pour l'acteur (individu, groupe, entreprise) ; *Internalisation* : *savoir-comportemental* relatif d'une part à la compétence (transformation de la connaissance en action) des ressources humaines qui matérialisent les qualités professionnelles de l'individu dans son espace de travail (l'environnement : le travail et ses outils de production), et au savoir-être (les qualités personnelles de l'individu) exprimé au travers du phénomène de l'intelligence émotionnelle.

## 2.2 Approche managériale du KM

Le fondement théorique de l'approche managériale du KM est la conception de l'entreprise comme un système ouvert (entrées/transformation/sorties) vis-à-vis de son environnement sur la base du concept d'enaction de Weick (1979). L'enaction décrit la relation entre entreprise et environnement comme deux sous-systèmes distincts, mais en interaction forte (confrontation), afin de maintenir une certaine stabilité dans la relation du système. Weick (1979) distingue un processus d'interaction par inclusion (l'un des deux systèmes doit dicter sa loi à l'autre) et un processus d'interaction par parallélisme (les systèmes négocient). Or, le concept d'enaction de Weick est important car il permet l'existence et l'opérabilité d'un système ouvert (entrées, sorties) mais aussi d'un système fermé (ni entrées, ni sorties) vis-à-vis de son environnement, et comme nous verrons plus loin l'approche de l'enaction de Maturana et Varela (1987) permet l'existence et l'opérabilité d'un système clos (organisation, structure, intelligence).

Les travaux de Ermine (1996), Tounkara (2002), Tounkara *et al.* (2002), Ermine (2003), Boughzala et Ermine (2004) sur le "patrimoine de connaissances" de l'entreprise (la connaissance est associée au métier) comme un système ouvert et les travaux de Grundstein (1996), Pachulsky *et al.* (2000), Pachulsky (2001), Grundstein et Rosenthal (2003), Ermine *et al.* (2006) sur les "connaissances cruciales" de l'entreprise (la connaissance est associée à l'action managériale) ont contribué à la caractérisation de l'approche managériale du KM sur la base du concept d'enaction de Weick (1979).

### 2.2.1 Paradigme de la marguerite

Si l'on fait l'hypothèse que l'entreprise (système ouvert) peut être décrite à travers un patrimoine de connaissances (approche de Ermine) ou un ensemble de connaissances cruciales (approche de Grundstein). Une "marguerite" (figure 2) permet de symboliser l'enaction environnement-patrimoine de connaissances (ou connaissances cruciales) à travers des processus internes (interaction par inclusion) et processus externes (interaction par parallélisme) (Ermine, 1996).

La marguerite est composée d'un cœur pour abriter le patrimoine de connaissances de l'entreprise, autour de celui-ci on a quatre pétales. Le cœur et les pétales sont définis dans une dynamique circulante permanente entre cinq processus (internes et externes) pour créer et gérer le patrimoine de connaissances. De gauche à droite, le processus externe de sélection par l'environnement (phase de projection) permet de repérer les connaissances métiers (cruciales) du business par sélection d'information (formulation des requêtes vers l'environnement externe) afin d'élaborer le corpus d'information. Le processus interne de capitalisation et de partage des connaissances (phase de renseignement ou d'intelligence) permet de préserver les connaissances métiers (cruciales) du business, et de le mettre à la disposition de tous les acteurs (individu, groupe, …) de l'entreprise. Le processus externe d'interaction avec l'environnement (phase de confrontation) permet de valoriser les connaissances métiers (cruciales) du business qu'on met en correspondance avec l'extérieur pour la détection du besoin. Le processus interne d'apprentissage et de création de connaissances (phase de création de sens) permet de faire évoluer le patrimoine de connaissances (connaissances cruciales). Le processus interne d'évaluation du patrimoine de connaissances (phase de développement) permet de mesurer la valeur du patrimoine de connaissances ou les connaissances cruciales de l'entreprise, par un système de mesure traduisant la rentabilité de leurs investissements en matière d'actifs immatériels.

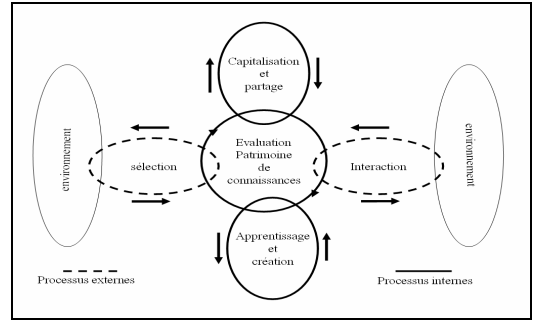

FIG. 2 – *Paradigme de la marguerite.*

Le maintien de la marguerite en vie se fait par une confrontation permanente avec son environnement au travers de la sélection d'information.

L'approche managériale du KM relie la connaissance au sujet, c'est-à-dire la connaissance est liée à l'action managériale (Grundstein parle de connaissances cruciales), ou bien la connaissance est reliée à l'objet, c'est-à-dire que la connaissance est liée au métier (Ermine parle de patrimoine de connaissances).

Gérer les connaissances de l'entreprise, selon l'enaction de Weick (1979), implique d'une part que l'entreprise et son environnement sont en confrontation permanente, et d'autre part, la mis en place des processus de sélection, de capitalisation et de partage, d'interaction, d'apprentissage, de création, et d'évaluation dans le but de repérer, préserver, valoriser, faire évoluer, et mesurer les connaissances métiers (cruciales) de l'entreprise, où la connaissance dans l'entreprise est décrite à travers un système ouvert (entrées, sorties).

### 2.3 Approche biologique du KM

Le fondement théorique de l'approche biologique du KM prend ses racines dans la conception de l'entreprise par rapport à une opération de distinction. Cette opération permet d'indiquer que les causes et les effets sont distinguables dans des espaces forts différents. L'un est le domaine conceptuel (l'organisation) du processus organisationnel (c'est un processus conceptuel de description abstrait de l'organisation). L'autre est le domaine physique (la structure) du processus structurel (c'est un processus physique de description matérielle de la structure, il s'agit bien ici de la description des propriétés des composants de la structure de l'organisation).

L'opération de distinction, est formulée par rapport à l'approche de l'enaction de Maturana et Varela (1987). L'enaction décrit la relation entre organisation et structure à partir de la spontanéité de trois processus : détermination structurelle, couplage structurel, et clôture opérationnelle. La détermination structurelle veut dire que l'entreprise afin de définir l'identité de l'organisation doit maintenir l'unité de la structure (l'entreprise assure une transformation définie dans et par l'organisation). Le couplage structurel signifie que l'entreprise pour maintenir son organisation doit modifier sa structure. La clôture opérationnelle se réfère au fait que l'entreprise est un système clos[6] au niveau de l'organisation (le résultat de la transformation se situe à l'intérieur des frontières du système lui-même), mais ouvert au niveau de la structure (le résultat de la transformation se situe à l'extérieur des frontières du système lui-même). Autrement dit, une fois définie l'organisation, il faut trouver la structure qui maintient l'identité dans et par l'organisation de l'unité. Comme Maturana et Varela (1987) soulignent « la clôture opérationnelle engendre une unité, qui à son tour spécifie un domaine phénoménal ». Et donc, il y a une spontanéité dans leur relation.

Partant de l'hypothèse que l'entreprise peut être approchée comme un système clos, c'est-à-dire, d'une part comme un système vivant caractérisé par l'identité (organisation) et l'unité (structure), et d'autre part comme un système viable caractérisé par l'autonomie (émergence d'un comportement intelligent[7]) nous formulons le paradigme de l'arbre.

### 2.3.1 Paradigme de l'arbre

Un "arbre" (figure 3) permet de symboliser l'enaction organisation-structure d'un organisme vivant et viable. L'arbre des connaissances[8] établi un rapport essentiel entre les processus de détermination structurelle, de couplage structurel, et de clôture opérationnelle qui ont lieu à l'intérieur de l'organisme de façon spontané afin de garantir l'auto-maintien de l'identité de l'organisation, l'auto-organisation de l'unité de la structure, et l'autogestion de l'autonomie au cours du temps, comme l'a dit Varela (1989) « tout système autonome est opérationnellement clos ».

La connaissance d'un point de vue biologique est « ce qui nous unit à tous les hommes de tous les temps et la manière par laquelle nous faisons apparaître en nous nos significations existentielles, la manière par laquelle celles-ci sont créées, stabilisées, transformées. C'est justement, dans ce processus d'apprentissage social qu'émerge en nous mêmes la signification du monde dans lequel nous vivons … la connaissance n'est pas armée comme un arbre avec un point de départ solide qui croît progressivement jusqu'à épuiser tout ce qu'il faut connaître, car la connaissance est un mécanisme "circulant" et "d'émergence de signification" … la connaissance est propre de l'être vivant, la connaissance est un grain que l'on sème dans le plus profond de nous mêmes » (Maturana et Varela, 1987). Autrement dit, la connaissance est une conduite (mécanismes et moyens d'agir) qui permet à l'organisme de faire seulement des choses qui n'affectent pas sa survie. La connaissance correspond au fait de faire émerger chez l'organisme un comportement intelligent. Et donc, l'enaction organisation-structure est un mécanisme "circulant" et "d'émergence de signification". Par conséquent, la connaissance ne doit pas obligatoirement impliquer des représentations vraies de la réalité objective.

L'approche biologique du KM relie alors la connaissance à la survie (processus d'organisation et de structuration du vivant) et à l'intelligence (processus d'émergence d'un comportement, une action). Le KM est un mécanisme de survie et d'intelligence plutôt que de construction de sens (problème de la signification, vraies représentations de la réalité).

---

[6] Système ouvert (entrées, sorties) et système fermé (ni entrées, ni sorties) différence vis-à-vis de son environnement. En revanche système clos (organisation, structure, intelligence) différence par rapport à une opération de distinction.

[7] Un système est "intelligent" s'il est capable, d'une part, d'être reconnu comme identité, et d'autre part, de maintenir l'unité de sa structure. Comme l'a si bien dit Varela (1989) « l'intelligence ne se définit plus comme la faculté de résoudre un problème mais comme celle de pénétrer un monde partagé ».

[8] Nous attirons ici l'attention sur le fait qu'il ne s'agit pas d'établir un arbre de compétences collectives pour l'entreprise comme celui développé par Authier et Lévy (1992).

Une approche sociotechnique pour le KM

Gérer les connaissances de l'entreprise, selon l'enaction de Maturana et Varela (1987), consiste à mettre en place des processus de détermination structurelle, de couplage structurel, et de clôture opérationnelle dans le but de créer, stabiliser et transformer les connaissances de l'entreprise, où la connaissance est décrite à travers un système clos (organisation, structure, intelligence) et non pas comme un système ouvert.

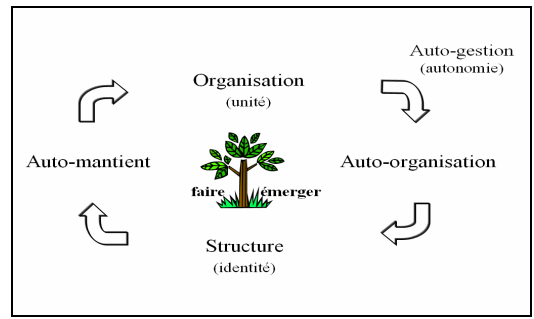

FIG. 3 – *Paradigme de l'arbre.*

Le KM symbolise la capacité de l'entreprise à capitaliser, partager, et créer de la connaissance afin de maintenir l'identité de l'organisation et l'unité de la structure comme un tout dans le temps.

### 2.4 Synthèses des approches

Le tableau 1 montre une synthèse des approches organisationnelle, managériale et biologique du KM selon l'aspect social.

Les approches organisationnelle, managériale et biologique du KM font apparaître de vrais domaines de recherche pour aboutir aux concepts, méthodes et outils autour de mécanismes génériques et processus du KM que nous avons synthétisés dans le tableau 1 :

- Comment créer des connaissances nouvelles et faire de l'apprentissage organisationnel dans l'entreprise ? En effet, selon l'approche organisationnelle du KM, le KM n'est pas un système de traitement de l'information mais bien un système de création des connaissances nouvelles et d'apprentissage organisationnel.

- Comment évaluer (mesurer) le savoir de l'entreprise ? En effet, selon l'approche managériale du KM, le KM permet d'évaluer les connaissances du business (patrimoine de connaissances ou connaissances cruciales) de l'entreprise par des méthodes et outils issus de l'expérience en terrain.

- Comment "faire-évoluer" (c'est-à-dire créer de nouvelles connaissances là, où il n'y a pas de savoir) et "faire-émerger" (c'est-à-dire créer des connaissances nouvelles à partir d'une représentation non symbolique de la réalité) les connaissances de l'entreprise ? En effet, le KM selon l'approche de l'enaction de Weick (mis en évidence par Grundstein (1996), Pachulsky (2001), Ermine (1996), Tounkara (2002)) permet la mise un place d'un système de connaissance bâti sur l'idée que la connaissance peut être définie au travers de l'information (données, traitements) qui prend une certaine signification (concepts, tâches) dans un contexte (domaine, activité) donné. En revanche, le KM selon l'approche de l'enaction de Maturana et Varela (mis en évidence par Limone et Bastias (2006), Jiménez (2005)) est propre d'un comportement intelligent.

|  | Approche Organisationnelle du KM | Approche Managériale du KM | Approche Biologique du KM |
|---|---|---|---|
| Analogie | ballon de rugby | marguerite | arbre |
| Fonctionnement | processus de causalité circulaire | sélection d'information | émergence d'un comportement intelligent |
| Mécanismes | créer, disséminer, valoriser | repérer, préserver, valoriser, faire évoluer, mesurer | créer, stabiliser, transformer |
| Processus | - Socialisation<br>- Extériorisation<br>- Combinaison<br>- Internalisation | - Sélection par l'environnement<br>- Capitalisation et partage<br>- Interaction avec l'environnement<br>- Apprentissage et création<br>- Evaluation | - Détermination structurelle<br>- Couplage structurel<br>- Clôture opérationnelle |

TAB. 1 – *Les approches du KM.*

## 3 Aspect technique du KM

En se plaçant maintenant dans l'aspect technique, le KM est vu à travers de tâches (processus et données) et de la technologie (méthodes et outils).

### 3.1 Approche ingénierie des connaissances et compétences du KM

La connaissance dans l'entreprise peut être considérée d'une part comme un objet (données) que l'on peut approcher à partir d'un projet d'ingénierie des connaissances (méthodes et outils du KM[9]), par exemple Ermine propose MASK (Method for Analyzing and Structuring Knowledge) et MKSM (Method for Knowledge System Management), permettant une analyse et une structuration d'un patrimoine de connaissance liée à la connaissance métier (une tâche en particulier) (Tounkara *et al*. (2002). Ermine (1996) affirme « MKSM est l'équivalent de MERISE pour les systèmes d'information, à savoir une méthode d'analyse de systèmes de connaissances pour aboutir à la conception d'un système opérationnel de gestion des connaissances », et d'autre part comme une action managériale (processus), par exemple Grundstein propose GAMETH (Global Analysis Methodology), une méthode pour le KM qui permet le repérage des connaissances cruciales de l'entreprise (Pachulsky *et al*. (2000). Grundstein (1996) dit « le management des connaissances … couvre toutes les actions managériales visant à actionner le cycle de capitalisation des connaissances afin de repérer, préserver, valoriser, transférer et partager les connaissances cruciales de l'entreprise ». La connaissance peut aussi être décrite à partir de cartes routières de compétences des acteurs de l'entreprise, par exemple Authier et Lévy (1992) proposent une méthode de repérage des savoirs et des savoir-faire, afin d'établir un arbre de compétences collectives pour l'entreprise. La méthode se trouve implémentée dans un progiciel appelé GINGO et développé par www.trivium.fr. Ces méthodes et autres[10] du KM symbolisent un macroscope (en empruntant le terme de Joël de Rosnay) pour observer l'acteur (individu, groupe, entreprise) dans son poste de travail (connaissance métier) en tant que facteur clé pour améliorer la capacité de l'entreprise pour maintenir l'organisation comme un tout dans le temps et non pas pour ajouter une autre application au parc informatique de l'entreprise, sinon nous risquons dans le futur d'enfermer le KM dans une problématique des systèmes à base de connaissances ou des systèmes experts.

---

[9] Depuis 2000 l'équipe de recherche ACACIA à l'INRIA Sophia Antiopolis, dirigée par Rose Dieng, a mis à disposition un catalogue pour le KM : *Méthodes et outils pour la gestion des connaissances* (publié en 2000), et *Méthodes et outils pour la gestion des connaissances : une approche pluridisciplinaire du Knowledge Management* (publié en 2001).

## 4   Conclusion

Nous avons présenté dans cet article, une approche sociotechnique du KM. Cette approche caractérise davantage l'aspect humain de la connaissance et son support technologique. C'est ainsi qu'ont été mises en évidence quatre perspectives, à savoir : l'approche organisationnelle de Nonaka et Takeuchi (fondée sur le concept de knowledge creating-company) ; l'approche biologique de Maturana et Varela (fondée sur le concept de l'enaction) ; l'approche managériale de Ermine (fondée sur le concept de la marguerite) ; et l'approche ingénierie des connaissances et compétences du KM.

Au carrefour de ces quatre approches, deux concepts sont le fondement de la problématique essentielle du KM. L'un est le concept de "faire-évoluer" les connaissances, c'est-à-dire de créer de nouvelles connaissances là, où il n'y a pas de savoir. L'autre est le concept de "faire-émerger" la connaissance, c'est-à-dire de créer des connaissances nouvelles à partir d'une représentation non symbolique de la réalité. En effet, tous les modèles de KM de nos jours sont gérés à partir du passé (les bonnes pratiques, le retour d'expérience, la communauté de pratiques, etc.) et non pas à partir de l'avenir (l'inconnu, le chaos, le désordre, etc.). Nous pensons que les concepts "faire-évoluer" et "faire-émerger" la connaissance sont fort intéressants pour réfléchir sur la question.

## Remerciments

## Références

---

[10] Plusieurs communautés autour du KM se sont installées : Club Gestion des Connaissances de Ermine (www.club-gc.asso.fr); O2 de Pachulsky (www.o2consulting.fr); et Arbre et Sens de Authier et Lévy (www.arbor-et-sens.org).

Une approche sociotechnique pour le KM

Weick K. (1979). *The Social Psychology of Organizing*. Addidon-Wesley.

# Summary

Following the theory of organizational, management and biological knowledge creation. This article takes a sociotechnical perspective on the organisational and technical issue of knowledge management.